\documentclass[letterpaper]{jpsj-suppl}
\usepackage{txfonts} 

\usepackage{url}

\title{Modern Dynamical Coupled-Channels Calculations for Extracting and Understanding the Nucleon Spectrum}

\author{Hiroyuki \textsc{Kamano}$^{1,2}$}

\inst{$^{1}$KEK Theory Center, Institute of Particle and Nuclear Studies (IPNS),
High Energy Accelerator Research Organization (KEK),
Tsukuba, Ibaraki 305-0801, Japan \\
$^{2}$J-PARC Branch, KEK Theory Center, IPNS, KEK, Tokai, Ibaraki 319-1106, Japan}

\email{kamano@post.kek.jp}

\recdate{}

\abst{
We give an overview of recent progress in the spectroscopic study of nucleon resonances
within the dynamical coupled-channels analysis of meson-production reactions.
The important role of multichannel reaction dynamics in understanding various properties 
of nucleon resonances is emphasized.
}

\kword{nucleon resonances, coupled-channels equations, multichannel unitarity, partial-wave analysis, meson-production reactions}

\begin{document}
\maketitle

\section{Introduction}

The spectroscopic study of nucleon resonances ($N^*$ and $\Delta^*$) dates back 
to the discovery of the $\Delta$ baryon by the Chicago University group in 1952~\cite{chicago}.
Here, the existence of a new baryon with the isospin $3/2$, 
which has come to be known as $\Delta(1232)3/2^+$, 
was suggested
from the rapid increase of the $\pi^+ p$ and $\pi^- p$ reaction total cross sections 
at $\sim 150$~MeV of the incident pion momentum and the ratios of the cross sections.
After 60 years of this discovery, nearly 50 $N^*$ and $\Delta^*$ baryons
have been reported, as listed by Particle Data Group~\cite{pdg14}.
However, as pointed out by the George Washington University group~(see the Introduction of 
Ref.~\cite{said06}), one still does not have any definitive conclusions 
for more than half number of the reported $N^*$ and $\Delta^*$ baryons, even for their existence.
The $N^*$ and $\Delta^*$ spectroscopy
therefore remains as a fundamental challenge in the hadron physics.

In the past, a number of static hadron models, such as constituent quark models~\cite{cqm} 
and models based on the Dyson-Schwinger equations~\cite{dse}, 
have been proposed to study the mass spectrum and quark-gluon substructure of hadrons.
In such static hadron models, the excited hadrons are usually treated as stable particles.
However, in reality, the excited hadrons strongly couple to the multihadron scattering states 
and can exist only as unstable resonances in hadron reactions. 
This fact raises an intriguing question how important
the dynamical effects arising from such a strong coupling to scattering states are
in understanding the mass spectrum, structure, and production mechanism 
of hadrons as resonant particles. 
To answer this question, the so-called dynamical coupled-channels (DCC) approaches
have been developed by a number of theoretical groups including us. 
These approaches have been applied to the analysis of various meson-production reactions
in the nucleon resonance region and have succeeded in providing new insight into 
dynamical contents of hadron resonances, 
which is difficult to be addressed by the static hadron models.
In this contribution, we give an overview of the DCC approaches
and present our recent efforts for the $N^*$ and 
$\Delta^*$ spectroscopy based on the so-called ANL-Osaka DCC approach~\cite{msl07,knls13,knls16}.

\section{$N^*$ and $\Delta^*$ spectroscopy: Physics of broad and overlapping resonances}
\label{sec:}

\begin{figure}[t]
\centering
\includegraphics[width=0.5\textwidth,clip]{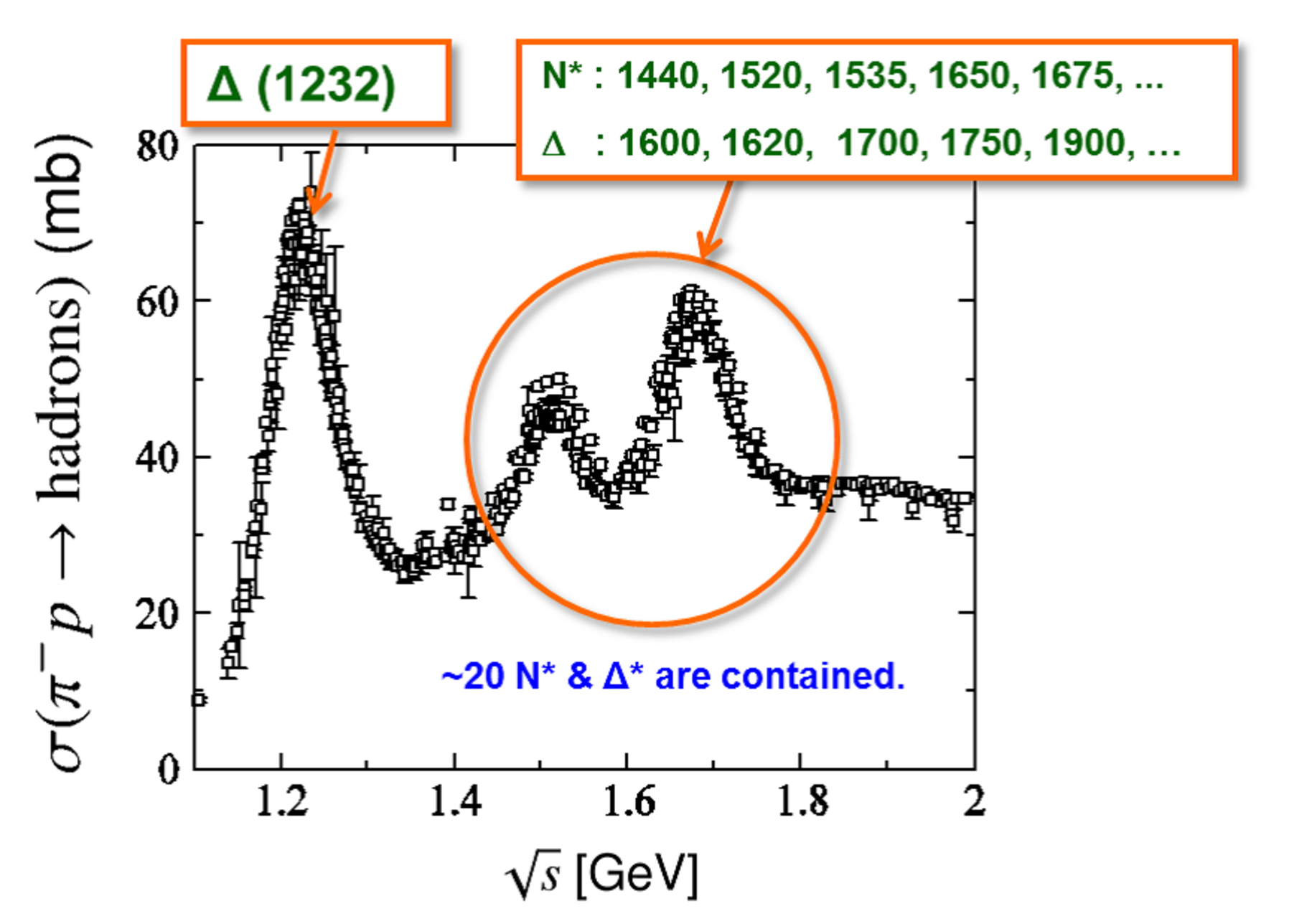}
\caption{\label{fig:pimptcs}
Total cross section for inclusive $\pi^- p$ reaction in the resonance region.
The first peak at $\sqrt{s}\sim 1.2$~GeV is produced solely by the $\Delta(1232)3/2^+$ resonance,
while the next two higher peaks contain $\sim 20$ $N^*$ and $\Delta^*$ resonances.
}
\end{figure}

The resonances usually appear as isolated peaks in the cross sections.
In fact, the first peak in the $\pi^- p$ reaction total cross section 
is attributed to the existence of the $\Delta(1232)3/2^+$ resonance (Fig.~\ref{fig:pimptcs}).
One then may expect that next two peaks at $\sqrt{s} \sim 1.5$~GeV and $\sqrt{s} \sim 1.7$~GeV 
are also produced by isolated resonances.
However, it is turned out that they contain $\sim 20$ $N^*$ and $\Delta^*$ resonances.
Furthermore, the decay widths of these resonances are found to be 
very broad, $\sim 300$~MeV on average, 
which can be even broader than the energy range of the two peaks.
This means that the $N^*$ and $\Delta^*$ resonances are highly overlapping 
with each other in energy, and thus a peak in the cross section does not necessarily 
mean the existence of an isolated resonance in the $N^*$ and $\Delta^*$ spectroscopy.
This situation is quite different from other systems such as heavy-quark hadrons,
atoms, and nuclei. 
In those systems, the resonances usually appear as clear and well-separated peaks 
in the cross sections.

The broad and overlapping nature of $N^*$ and $\Delta^*$ resonances makes 
their experimental identification very difficult.
Cooperative works between experiments and theoretical analyses 
are therefore indispensable for the $N^*$ and $\Delta^*$ spectroscopy.
In fact, tremendous efforts in such a direction have been performed since the late 90s.
A huge amount of high statistics data of meson-production reactions off the nucleon 
were obtained from photon- and electron-beam facilities, 
such as ELPH, ELSA, JLab, MAMI, and SPring-8,
and were brought to theoretical analysis groups using coupled-channels approaches
such as ANL-Osaka, Bonn-Gatchina, J\"uelich, and SAID~\cite{nstar2015}.
The analysis groups then performed comprehensive partial-wave analyses of the data and 
extracted various properties of $N^*$ and $\Delta^*$ resonances
defined by poles of scattering amplitudes in the complex-energy plane.
In parallel with this, the analysis groups gave feedback about what data are 
further needed for more complete determination of $N^*$ and $\Delta^*$ resonances.
With this close cooperation between experiments and theoretical analyses, 
significant progress has been achieved for the $N^*$ and $\Delta^*$ spectroscopy in recent years.

\section{Multichannel unitarity and dynamical coupled-channels approaches}

\begin{figure}[t]
\centering
\includegraphics[width=0.5\textwidth,clip]{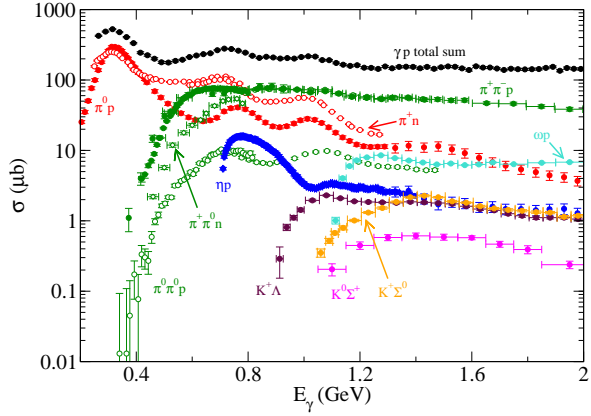}
\caption{\label{fig:gptcs}
Total cross section for exclusive $\gamma p$ reactions in the resonance region.
}
\end{figure}

The unitarity of the multichannel $S$-matrix, $S^\dag S = 1$, is the key to performing 
the coupled-channels analysis 
and making reliable extraction of $N^*$ and $\Delta^*$ resonances from reaction data.
Defining the $T$-matrix as $S_{ba}= \delta_{ba} -i2\pi \delta(E_b-E_a)T_{ba}$, 
where $E_a = \sum_i E_{a,i}(\vec p_{a,i})$ with $E_{a,i}$ and $\vec p_{a,i}$ being the energy and momentum of 
the $i$th particle belonging to the channel $a$, respectively.
The unitarity $S^\dag S = 1$ gives 
the following condition obeyed by the on-shell $T$-matrix elements~\cite{gw} (the generalized optical theorem):
\begin{equation}
T_{ba}(E) - T^\dag_{ba}(E) = -2\pi i \sum_c T^\dag_{bc}(E)\delta(E-E_c)T_{ca}(E),
\label{eq:opt}
\end{equation}
where the subscripts represent the reaction channels, and $E=E_a=E_b$.

There are two critical reasons why the multichannel unitarity is so important.
First, it ensures the conservation of probabilities in multichannel reactions.
As can be seen from the $\gamma p$ reaction total cross sections presented in Fig.~\ref{fig:gptcs}, 
many inelastic channels open in the resonance region. 
It is almost impossible to treat all of the inelastic reactions 
consistently in a single reaction framework unless 
the transition probabilities are automatically conserved by the multichannel unitarity.
Second, the multichannel unitarity condition [Eq.~(\ref{eq:opt})] properly defines 
the analytic structure (branch points and unitarity cuts, etc.)
of the scattering amplitudes in the complex-energy plane.
Any reaction framework that does not satisfy this condition would fail 
to make a proper analytic continuation of the amplitudes, and this 
may result in picking up wrong signals of resonances.

It is known that Eq.~(\ref{eq:opt}) is satisfied by any $T$-matrix given by the Heitler equation~\cite{gw}:
\begin{equation}
T_{ba}(E) = K_{ba}(E) + \sum_c K_{bc}(E)[-i\pi \delta(E-E_c)]T_{ca}(E),
\label{eq:heitler}
\end{equation}
where $K_{ba}(E)$ is known as the (on-shell) $K$-matrix, and 
the unitarity condition requires this to be Hermitian for real $E$.
Since the unitarity condition does not give any further constraints on the form of $K$-matrix as a function of $E$, 
usually two approaches are taken for parametrizing the $K$-matrix.
One is called the (on-shell) $K$-matrix approach, where the $K$-matrix is simply parametrized 
as a sum of polynomials and pole terms of $E$.
In this case, the Heitler equation can be reduced to a simple algebraic equation 
at least for the case of two-body reactions.
Another is called the dynamical-model approach, in which the $K$-matrix is obtained 
by solving the following equation:
\begin{equation}
K_{ba}(E) \equiv K_{ba}(\vec p_b,\vec p_a;E)
= V_{ba}(\vec p_b,\vec p_a;E)
+ {\sum_d}' {\cal P}\int d\vec p_d V_{bd} (\vec p_b,\vec p_d;E)\frac{1}{E-E_d} K_{da}(\vec p_d,\vec p_a;E),
\label{eq:k-dcc}
\end{equation}
where $V$ is the transition potential defined by some model Hamiltonian; 
$\vec p_a$ symbolically denotes the momenta of all particles in the channel $a$, $\vec p_a = (p_{a,1},..,p_{a,N_a})$
with $N_a$ being number of the particles in the channel $a$;
the symbol ${\cal P}$ means taking the Cauchy principal value for the integral over the momentum variable $\vec p_d$;
and the symbol ${\sum'_d}$ means taking summation or integral 
for all variables of the channel $d$ except for the momenta.
The second term in the right hand side of Eq.~(\ref{eq:k-dcc})
describes the off-shell rescattering effect in the reaction processes.
The Heitler equation~(\ref{eq:heitler}) combined with the $K$-matrix given by Eq.~(\ref{eq:k-dcc})
is nothing but the Lippmann-Schwinger integral equation describing the quantum scattering.
Our approach (the ANL-Osaka DCC approach) belongs to the dynamical-model approach.

The (on-shell) $K$-matrix approach seems more ``economical'' than the dynamical-model approach
in terms of numerical analysis of reaction data.
In fact, the numerical cost of the (on-shell) $K$-matrix approach is basically much cheaper 
than the dynamical-model approach, because in the latter case, 
one has to solve the very time-consuming off-shell integral equation.
In addition, the (on-shell) $K$-matrix approach is much easier to obtain 
a good fit to the data because one can parametrize the $K$-matrix as one likes.
On the other hand, in the dynamical-model approach, 
the form of the $K$-matrix, which is given from the potential $V$ by Eq.~(\ref{eq:k-dcc}), 
is severely constrained by a model Hamiltonian employed as a theoretical input.
Therefore, the (on-shell) $K$-matrix approach would be enough 
if enough amounts of precise data are available and if
what one wants to know is just the resonance pole positions and residues 
of the on-shell scattering amplitudes.
However, if one further wants to understand the physics of reaction dynamics 
behind various properties of hadron resonances, then 
the dynamical-model approach is necessary, 
because such a study can be achieved only by appropriately modeling 
the reaction processes and solving a proper quantum scattering equation.
This is why we employ the dynamical-model approach.

\begin{figure}[t]
\centering
\includegraphics[width=0.65\textwidth,clip]{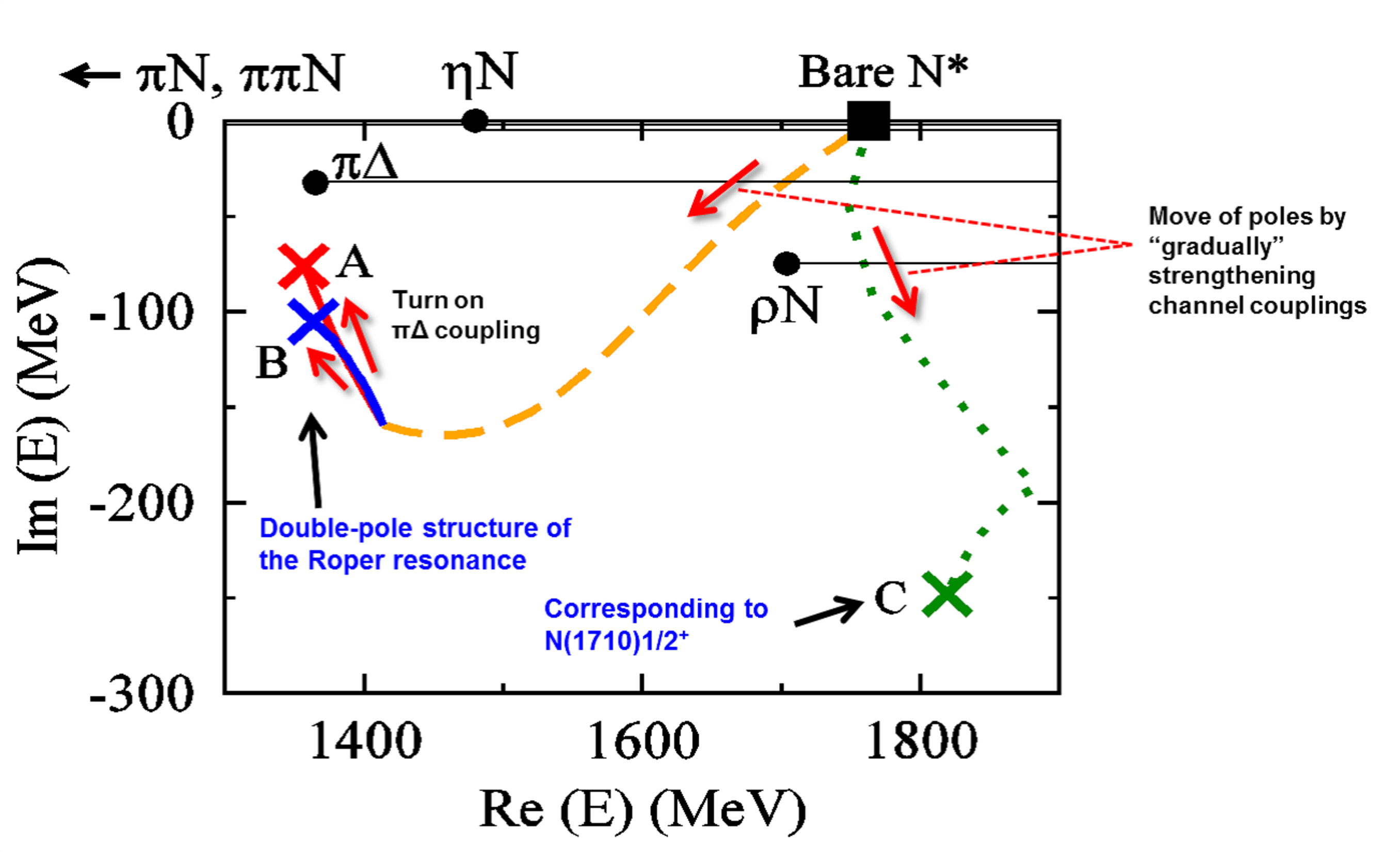}
\caption{\label{fig:p11}
Dynamical origin of $P_{11}$ ($J^P = 1/2^+$) $N^*$ resonance poles~\cite{sjklms10}
within the dynamical coupled-channels model developed in Ref.~\cite{jlms07}.
Poles A and B are the double-pole (pole and shadow-pole) structure of the Roper resonance
with respect to the $\pi \Delta$ channel, while pole C corresponds to $N(1710)1/2^+$.
Filled square is the so-called ``bare'' $N^*$ state, which is defined as an eigenstate of 
the model Hamiltonian for which the coupling to the reaction channels are turned off.
Dynamical effects originating from multichannel reaction processes
trigger the generation of all three resonance poles A, B, and C from the single bare $N^*$ state.
See Ref.~\cite{sjklms10} for the details of the description of the figure.
}
\end{figure}

Let us present two examples that clearly show an importance of 
using dynamical-model approaches to clarify the role of 
reaction dynamics in understanding properties of $N^*$ and $\Delta^*$ resonances.
One is the dynamical origin of $P_{11}$ $N^*$ resonances~\cite{sjklms10}.
Figure~\ref{fig:p11} shows pole positions of $P_{11}$ $N^*$ resonances
extracted from a dynamical model developed in Ref.~\cite{jlms07}.
Here, the poles $A$ and $B$ are well known as the double-pole (pole and shadow-pole~\cite{eden}) 
structure of the Roper resonance with respect to the $\pi \Delta$ channel,
which has been observed also in Refs.~\cite{arndt85,cw90,said06,doring09}
and mentioned by PDG~\cite{pdg14},
while the pole $C$ corresponds to the $N^*(1710)1/2^+$ resonance.
On the other hand, the so-called ``bare'' $N^*$ state, which has the real 
mass of 1763 MeV (the filled square in Fig.~\ref{fig:p11}), 
is the one defined as an eigenstate of the model Hamiltonian for which 
the coupling to the reaction channels are turned off.
The bare $N^*$ state therefore conceptually corresponds to 
a baryon state obtained in the static hadron models.
Then it was found that within the model developed in Ref.~\cite{jlms07}, 
all of the presented three $P_{11}$ resonance poles (the poles A, B, C) are generated from 
this single bare state as a result of its coupling to the multireaction channels~\cite{eden}.
This implies that a na\"ive one-to-one correspondence between 
the physical resonances and the baryons within static hadron models, within which
the dynamical effects originating from coupling to the reaction channels are neglected, 
does not exist in general. 
Furthermore, the reaction dynamics can produce a sizable mass shift, 
as can be seen from the mass difference between the bare state and the Roper resonance.
These findings for the $P_{11}$ resonance mass spectrum might be still dependent 
on this particular model, and further investigations combined with other quantities 
such as electromagnetic transition form factors would be necessary 
to obtain more conclusive results. 
However, at least one can say that the mass spectrum of physical resonances can be 
very different from that obtained in static hadron models, and 
one cannot neglect reaction dynamics in understanding the nucleon resonances.

\begin{figure}[t]
\centering
\includegraphics[width=0.75\textwidth,clip]{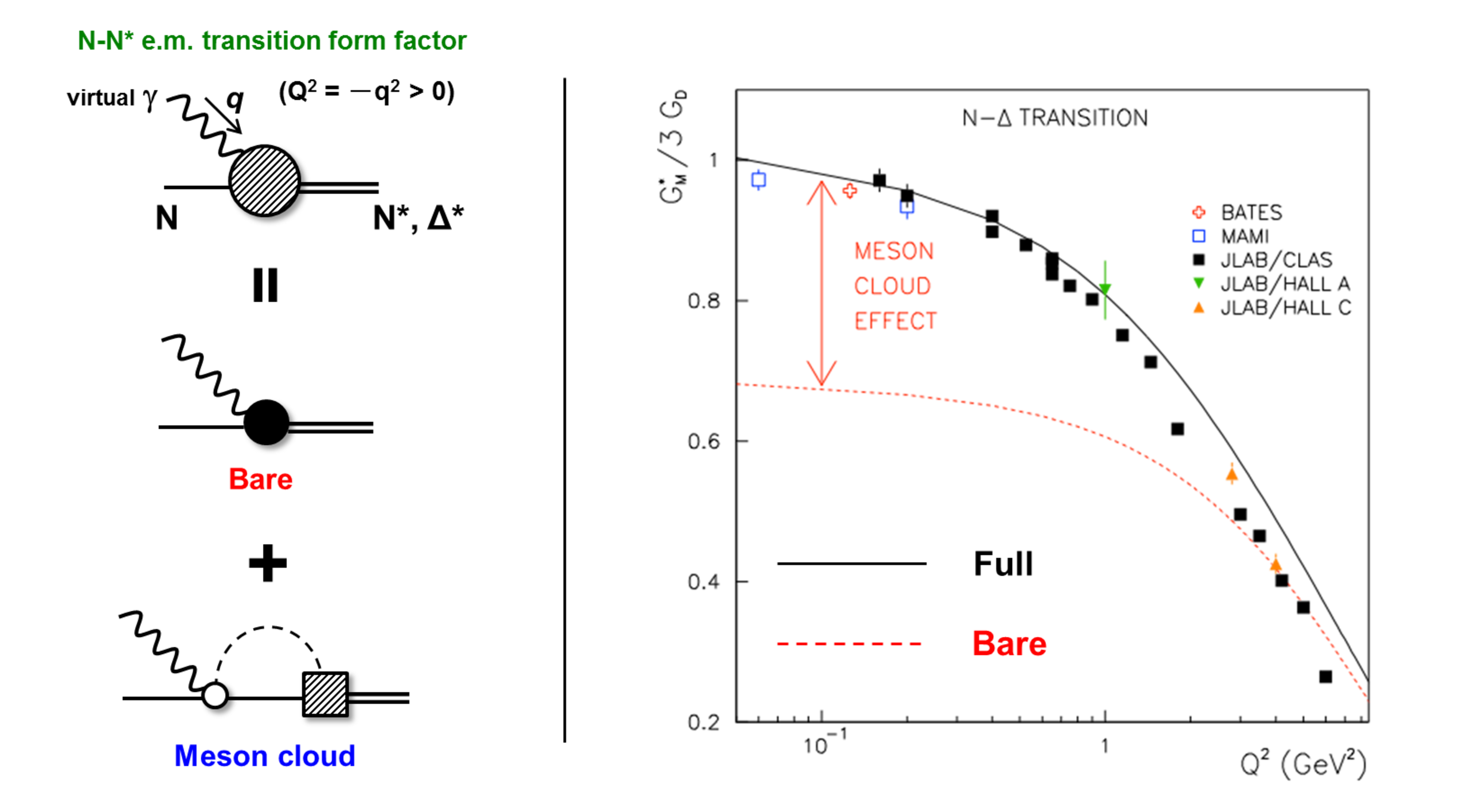}
\caption{\label{fig:n-d-emff}
(Left)
Schematic diagram of the electromagnetic form factor for the transition from 
the nucleon to a nucleon resonance.
Within the dynamical-model approach, the form factor is given by a sum of
the bare $N^*$ and meson-cloud contributions.
(Right)
$Q^2$ dependence of the $M1$ transition form factor,  $G_M^*(Q^2)$,
between the nucleon and the $\Delta(1232)3/2^+$ resonance,
divided by $3 G_D(Q^2)$ with $G_D(Q^2)=(1+Q^2/[0.71(\textrm{GeV/c})^2])^{-2}$.
The solid (dashed) curve is the results of the full dressed (bare) form factor.
The result is given from a dynamical model developed in Ref.~\cite{sl2}.
}
\end{figure}

Another example indicating the importance of using dynamical-model approaches is
the electromagnetic transition form factors between the nucleon and 
nucleon resonances probed by the virtual photon (the left side of Fig.~\ref{fig:n-d-emff}).
Here, $Q^2$ defined by $Q^2\equiv -q^2$ with $q$ being the four-momentum of virtual photon 
represents the ``resolution'' of the virtual photon, 
and hence the $Q^2$ dependence of the form factors is expected 
to provide crucial information on the substructure of the $N^*$ and $\Delta^*$ resonances.
Because of this, the electromagnetic transition form factors are being actively investigated 
both experimentally and theoretically, and this has opened a great opportunity 
to make a quantitative study of the substructure of the $N^*$ and $\Delta^*$ resonances 
in close relation with experimental data (see, e.g., Ref.~\cite{az13}).
The right side of Fig.~\ref{fig:n-d-emff} shows the $M1$ transition 
form factor between the nucleon and 
the $\Delta(1232)3/2^+$ resonance extracted from a dynamical model developed in Ref.~\cite{sl2}.
Within dynamical models, the full dressed form factor consists of 
the bare form factor and the meson cloud, where
the latter purely originates from the reaction dynamics. 
It is found that $\sim 30$ \% of the full dressed form factor 
comes from the meson cloud in the low $Q^2$ region.
It is notable that most of the available static hadron models, 
in which the reaction dynamics is not taken into account, 
indeed give the form factor close to the bare bare form factor, 
but not to the full dressed form factor.
One can also observe from the right side of Fig.~\ref{fig:n-d-emff} 
that the meson cloud effect becomes smaller as $Q^2$ increases.
These results obtained from the dynamical-model approach suggests that 
at a long distance scale the $\Delta(1232)3/2^+$ resonance can be understood 
as a constituent quark-gluon core surrounded by dense meson clouds, and 
the core part gradually emerges at shorter distance scales.
To obtain deeper insight into the transition form factors in the high $Q^2$ region, 
in which the contribution of quark-gluon core is expected to dominate, 
experimental determination of the transition form factors through 
the measurement of electroproduction reactions in the region of
$5\lesssim Q^2 \lesssim 12$ GeV$^2$ is planned at CLAS12~\cite{clas12-proposals}.

\section{Recent results from ANL-Osaka DCC analysis}

Now let us move on to presenting our recent efforts for the $N^*$ and $\Delta^*$ spectroscopy 
based on the ANL-Osaka DCC model~\cite{msl07,knls13,knls16}.
The basic formula of the model is the multichannel 
Lippmann-Schwinger equation obeyed by the partial-wave amplitudes:
\begin{equation}
T^{(J^P I)}_{b,a} (p_b,p_a;E) = 
V^{(J^PI)}_{b,a} (p_b,p_a;E)
+\sum_c \int_C dp_c p_c^2 V^{(J^PI)}_{b,c} (p_b,p_c;E) G_c(q;E) T^{(J^PI)}_{c,a} (p_c,p_a;E),
\label{lseq}
\end{equation}
where the subscripts represent the reaction channels and their spin and angular momentum quantum numbers;
$p_a$ represents the magnitude of the relative momentum of the channel $a$ in the center-of-mass system;
and $(J^P I)$ specifies the total angular momentum, parity and total isospin of the considered
partial wave.
At present, we have taken into account the $\pi N$, $\eta N$, $\pi \Delta$, $\rho N$, $\sigma N$, $K\Lambda$, and $K\Sigma$ channels,
where the $\pi\Delta$, $\rho N$, and $\sigma N$ are 
the quasi-two body channels that subsequently decay into the three-body $\pi\pi N$ channel.
The Green's function $G_c(q;E)$ is given by
$G_c(q;E)=1/[E-E_M(q)-E_B(q)+i\varepsilon]$ for $c= \pi N, \eta N, K\Lambda, K\Sigma$,
while $G_c(q;E)=1/[E-E_M(q)-E_B(q)-\Sigma_c(q;E)]$ for $c =\pi\Delta, \rho N, \sigma N$,
where $M$ and $B$ are the meson and baryon contained in the channel $c$, 
$E_M (q) = (m_M^2 + q^2)^{1/2}$ is the energy of the particle $M$,
and $\Sigma_c(q;E)$ is the self energy of $\Delta$, $\rho$, or $\sigma$ in the presence of the spectator particle.
For the $\pi\Delta$, $\rho N$, and $\sigma N$ channels, the Green's function produces the three-body cut due to the
opening of the $\pi \pi N$ channel in the intermediate reaction processes.

Our physics input is contained in the transition potential.
In our framework, the potential consists of three pieces: 
\begin{equation}
V^{(J^PI)}_{b,a} (p_b,p_a;E) = v^{(J^PI)}_{b,a} (p_b,p_a;E) + Z^{(J^PI)}_{b,a} (p_b,p_a;E) + 
\sum_{N^*_n}\frac{ \Gamma_{b,N_n^*}(p_b) \Gamma_{N_n^*,a}(p_a)} {E - M^0_{N_n^*}}.
\label{pot}
\end{equation}
The first two terms describe the so-called non-resonant processes including only 
the ground state mesons and baryons belonging to each flavor SU(3) multiplet, 
and the third term describes the propagation of the bare $N^*$ states.
We quote Ref.~\cite{knls13} for the details of the potential.
It is noted that the $Z$-diagram potential [the second term of Eq.~(\ref{pot})] 
also produces the three-body $\pi \pi N$ unitarity cut, 
and the implementation of both the $Z$-diagram potential 
and the self-energy in the Green's functions 
is necessary for maintaining the three-body unitarity.
Within our framework, the bare $N^*$ states are defined as eigenstates of the Hamiltonian 
for which the couplings to the reaction channels are turned off.
So by definition, our bare $N^*$ states would correspond to the hadron states 
obtained from the static hadron models such as constituent quark models.
By solving Eq.~(\ref{lseq}), the bare $N^*$ states couple to the reaction channels considered,
and then they get complex mass shifts and become resonance states.
Of course there is another possibility that the hadron-exchange potential 
[the first and second terms of Eq.~(\ref{pot})]
generates resonance poles dynamically.
Our model contains both possibilities.

\begin{figure}[t]
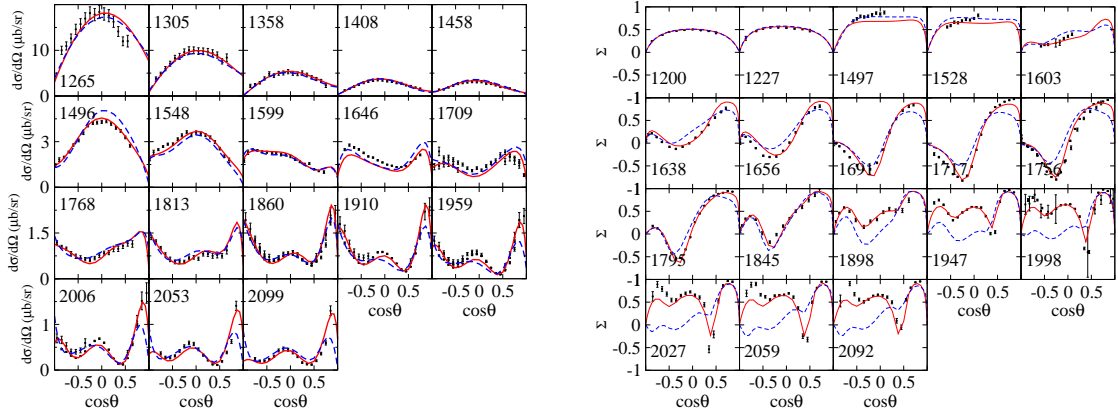

\centering
\includegraphics[width=0.45\textwidth,clip]{gppi0p-dc}
\qquad
\includegraphics[width=0.45\textwidth,clip]{gppi0p-s}
\caption{\label{fig:gppi0p}
(Left) Differential cross section $d\sigma/d\Omega$ for $\gamma p \to \pi^0 p$
(Copyright 2016 The Physical Society of Japan~\cite{nstar-kamano}).
(Right) Photon asymmetry $\Sigma$ for $\gamma p \to \pi^0 p$.
The numbers shown in each panel are the corresponding total scattering energy $W$ in MeV. 
The red solid (blue dashed) curves are the results from 
our original 8-channel model~\cite{knls13} (its latest updated model~\cite{knls16}).
See Ref.~\cite{knls13} for the references of the data.
}
\end{figure}

To study $N^*$ and $\Delta^*$ resonances, we first need to determine the model parameters such as coupling constants and bare baryon masses, and this is done by fitting to the data of meson production reactions.
Our latest 8-channel model developed and updated in Refs.~\cite{knls13,knls16} was constructed by 
a simultaneous fit of more than 27,000 data points of the differential cross sections and 
spin polarization observables for $\pi N\to \pi N$ up to $W=2.3$ GeV, 
$\pi N\to \eta N, K\Lambda, K\Sigma$ and $\gamma p\to \pi N \eta N, K\Lambda, K\Sigma$ up to $W=2.1$ GeV, 
and $\gamma \textrm{`}n\textrm{'} \to \pi N$ up to $W = 2$ GeV.
As an example of our fit, the differential cross section and photon asymmetry 
for the $\gamma p \to \pi^0 p$ reaction
are presented in Fig.~\ref{fig:gppi0p}.
Here the results from our original 8-channel model developed 
in 2013~\cite{knls13} are compared with
the latest updated version~\cite{knls16},
showing visible improvements of our fit 
at several energies, particularly for the photon asymmetry.

\begin{figure}[t]
\centering
\includegraphics[width=0.8\textwidth,clip]{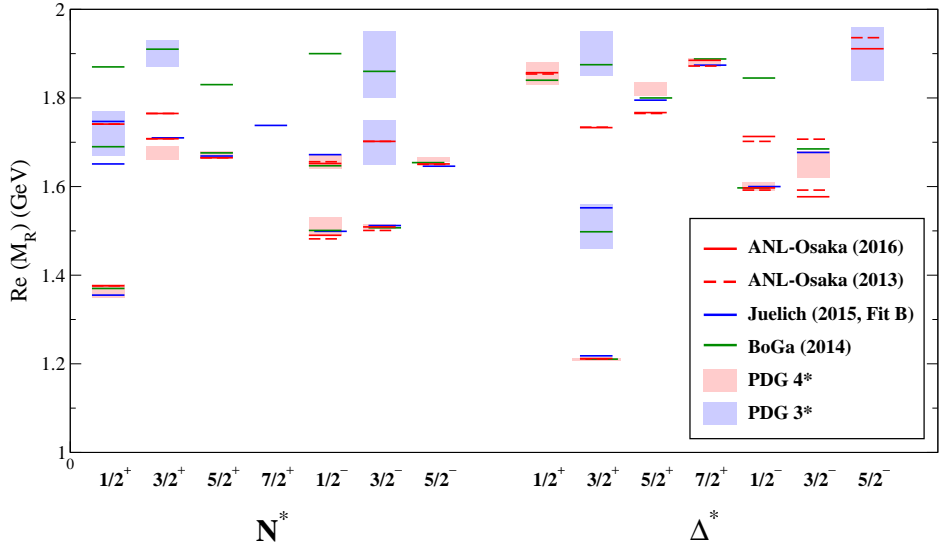}
\caption{\label{fig:spectrum}
Mass spectra for $N^*$ and $\Delta^*$ resonances. 
Real parts of the resonance pole masses $M_R$ are plotted. 
The red solid~\cite{knls16} and dashed~\cite{knls13} lines are the results from the ANL-Osaka analyses, 
while the blue and green solid lines are the results from the J\"ulich~(Model B of \cite{juelich}) and Bonn-Gatchina~\cite{bg} analyses.
The spectra of four- and three-star resonances rated by PDG~\cite{pdg14} are also presented with the red and blue filled squares,
respectively, which represent the range of the resonance masses assigned by PDG.
Here we present only the resonances that have the decay width less than 400 MeV.
}
\end{figure}

Figure~\ref{fig:spectrum} shows the real parts of the extracted resonance pole masses.
Here, our results from Refs.~\cite{knls13,knls16} are compared with
those obtained from the other coupled-channels analyses by the J\"ulich~\cite{juelich} and Bonn-Gatchina~\cite{bg} groups.
One can see that the existence and mass values agree very well for the lowest states in most spin-party states.
Actually, the community has now more or less arrived at a consensus that 
the existence and mass spectrum for low-lying $N^*$ and $\Delta^*$ resonances below $\textrm{Re}(M_R) \sim 1.7$~GeV 
has been firmly established.
One exception is the second $P_{33}$ resonance, the Roper-like state of the $\Delta$ baryon.
Although its existence is fairly established, the value of the pole mass is fluctuated a lot 
between the coupled-channels analyses.
In fact, our results appear much higher than the J\"ulich and Bonn-Gatchina results.
A major reason for this is because this resonance couples weakly to the $\pi N$ and $\gamma N$ channels 
and thus it is hard to establish the resonance with the single $\pi$ production data.
However, we find that this resonance has a large decay branching ratio to the three-body $\pi \pi N$ channel~(see, e.g., Fig.~6 of Ref.~\cite{e45}).
This implies that the $\pi \pi N$ production data are expected to provide crucial information on establishing 
the second $P_{33}$ resonance.
In this regard, the J-PARC E45 experiment~\cite{e45}, 
in which the high statistics measurement of the $\pi^\pm p \to \pi \pi N$ reactions will be performed,
is very promising to resolve this issue because 
only $I=3/2$ $\Delta$ resonances selectively appear in the direct $s$-channel process
for the case of $\pi^+ p$ reactions. 
We will improve our DCC model, which is ready for computing 
observables of the $\pi N \to \pi \pi N$ reactions~\cite{pi2pi1,pi2pi2},
once the new data are available from the J-PARC E45 experiment~\cite{e45}.

\begin{figure}[t]
\centering
\includegraphics[width=0.8\textwidth,clip]{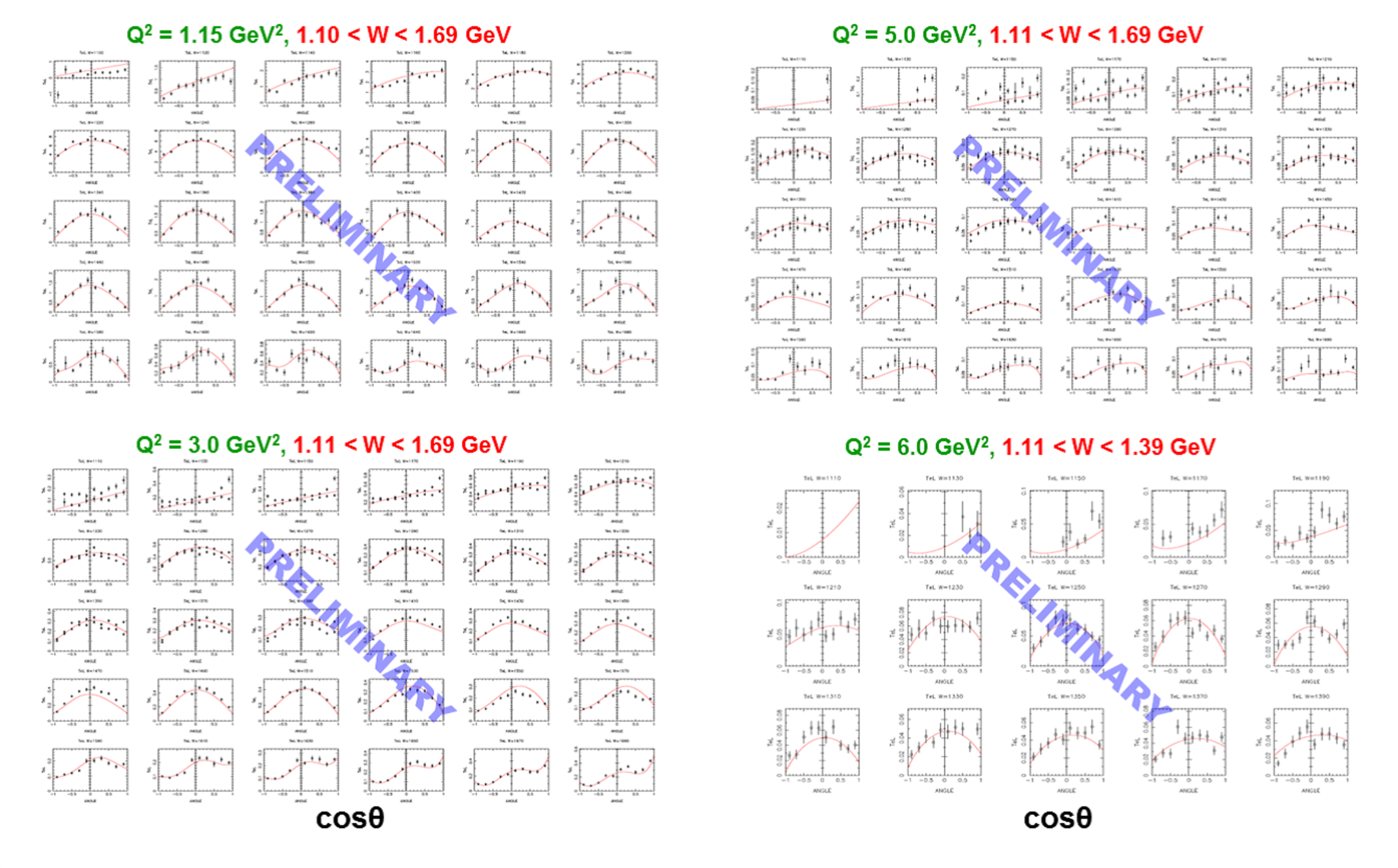}
\caption{\label{fig:e1pi}
Preliminary results for the analysis of $p(e,e'\pi^0)p$ reaction within the ANL-Osaka DCC model.
The results are presented for the structure function $\sigma_T + \varepsilon \sigma_L$
at $Q^2=1.15, 3, 5, 6$ GeV$^2$.
The data for the structure functions are the ones extracted in Ref.~\cite{joo-smith} from
the $p(e,e'\pi^0)p$ reaction cross sections published by CLAS~\cite{clas}.
}
\end{figure}

We also put effort into the analysis of the available data of electroproduction reactions 
to determine the electromagnetic transition form factors between the nucleon and nucleon resonances.
We currently focus on analyzing the single pion electroproductions 
data from CLAS in the kinematical region up to $Q^2 = 6$~GeV$^2$ and $W=1.7$~GeV.
Figure~\ref{fig:e1pi} shows some preliminary results of our ongoing analysis.
In the analysis, we use the so-called structure functions as the data to analyze, 
which were installed in our analysis with the help of K.~Joo and L.~C.~Smith~\cite{joo-smith}.
We see that our current results reproduce the data reasonably well up to $Q^2 = 6$ GeV$^2$.

\begin{figure}[t]
\centering
\includegraphics[width=0.75\textwidth,clip]{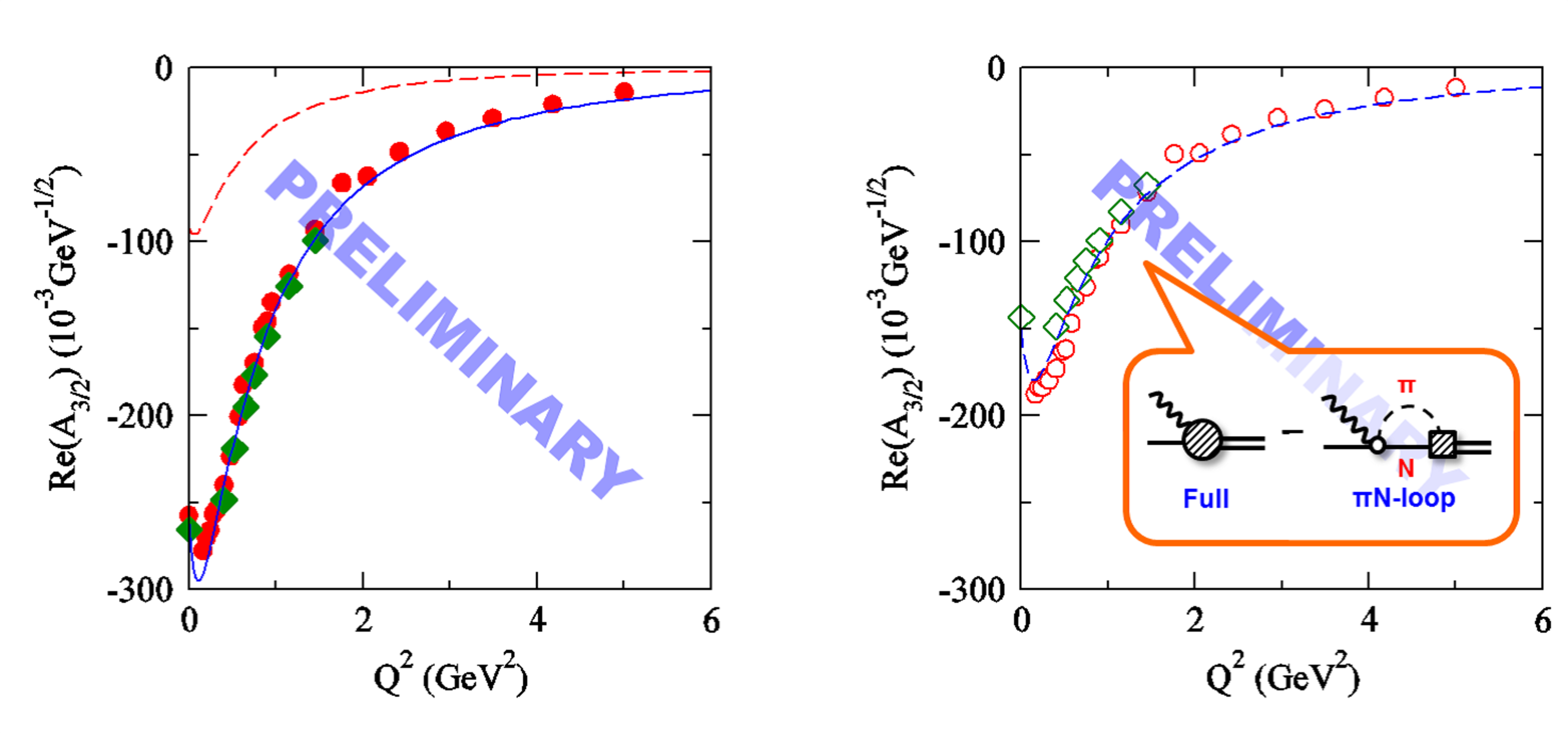}
\caption{\label{fig:n-d-emff-2}
Real part of the helicity amplitude $A_{3/2}$ for the transition between the nucleon and the 
$\Delta(1232)3/2^+$ resonance evaluated at the pole position.
(Left panel): The full form factors from the current ANL-Osaka DCC analysis (red filled circles),
our previous analysis~\cite{jklmss09,ssl2} (green filled diamonds),
and the Sato-Lee model~\cite{sl2} (blue solid curve).
The red dashed curve is the meson-cloud contribution from the current ANL-Osaka DCC analysis.
(Right panel): The form factors without pion-cloud ($\pi N$-loop) contribution. 
The results are from the current ANL-Osaka DCC analysis (red circles),
our previous analysis~\cite{jklmss09,ssl2} (green diamonds),
and the Sato-Lee model~\cite{sl2} (blue dashed curve).
}
\end{figure}

Figure~\ref{fig:n-d-emff-2} shows the real parts of the $A_{3/2}$ helicity amplitudes 
for the electromagnetic transition from the nucleon to the 
$\Delta(1232)3/2^+$ resonance evaluated at the pole position.
In the left panel, the full result extracted from our current analysis based on the ANL-Osaka 
DCC model is presented by the red circles, 
while its meson cloud contribution is plotted in the red dashed curve.
Comparing with Fig.~\ref{fig:n-d-emff}, one can see again that the meson-cloud contribution 
is almost 30\% at low $Q^2$ region, and its percentage becomes smaller as $Q^2$ increases.
In the same panel, the results from our previous analysis~\cite{jklmss09,ssl2} 
and from the Sato-Lee model~\cite{sl2} are also presented.
One can clearly see that all of the three results agree very well with each other, 
indicating that the transition form factors associated with this first $P_{33}$ resonance 
has been firmly established.
The right panel shows a comparison of the form factors 
without the pion-cloud contribution, which is defined as the full dressed form factor 
from which only the $\pi N$-loop contribution is subtracted.
These results are also found to agree well between the three models, 
even though their dynamical contents are rather different~\cite{footnote}.
This result would be quite remarkable because in general the separation of 
the bare and meson-cloud contributions is dependent on models, but this agreement implies 
that pion-cloud contributions might be nearly independent of the dynamical models employed.
To arrive at a more definitive conclusion on this interesting finding, however, we have to 
make further investigations of the transition form factors including other resonances as well, 
and this is in progress.

\section{Summary}
\label{sec:summary}

We have presented recent efforts for the spectroscopic study of $N^*$ and $\Delta^*$ resonances 
based on the dynamical coupled-channels approach.
The $N^*$ and $\Delta^*$ baryons are the system of very broad and highly 
overlapping resonances, and this requires close cooperative works 
between the experiments and the theoretical analyses with coupled-channels approaches
to accomplish a reliable extraction of $N^*$ and $\Delta^*$ resonances from reaction data. 
Tremendous efforts for this cooperation have led to a recent significant progress of 
the $N^*$ and $\Delta^*$ spectroscopy.
Because of the broad and overlapping nature of the $N^*$ and $\Delta^*$ resonances, 
the reaction dynamics plays a crucial role in understanding the physics behind 
their spectrum and substructure. 
We have shown that the dynamical coupled-channels approach is one of 
the most suitable approaches to study it, as demonstrated with 
the dynamical origin of $P_{11}$ $N^*$ resonances and the meson-cloud effect 
in the electromagnetic transition form factors. 
The existence and mass values for the low-lying nucleon resonances have now been 
firmly established (with one exception of the second $P_{33}$ resonance). 
The next important task is therefore to establish the spectrum of high-mass resonances.
In this regard, the so-called (over-)complete experiments for meson photoproduction 
reactions are underway at photon- and electron-beam facilities such as ELSA, JLab, and MAMI, 
and new high statistics data are continuously published.
Another major topic in the $N^*$ and $\Delta^*$ spectroscopy is to determine the $Q^2$ 
dependence of electromagnetic transition form factors for the well-established 
low-lying resonances. 
In this regard, huge amount of electroproduction data are being published from CLAS and 
new experiments are planned at CLAS12~\cite{clas12-proposals}.
To contribute to these interesting topics, 
the extension of our coupled-channels approach by including more reaction channels is underway.

Finally, the theoretical framework of the ANL-Osaka DCC approach itself is quite general, 
and it has been applied also to the spectroscopy of $S = -1$ hyperon resonances via
the comprehensive analysis of $K^- p$~\cite{knlskp1,knlskp2} and $K^- d$~\cite{kl16} reactions,
the meson spectroscopy via the analysis of three-meson decay processes~\cite{meson1,meson2},
and the neutrino-induced reactions~\cite{neutrino1,neutrino2}
associated with the neutrino-oscillation experiments in the multi-GeV region.
More efforts will be put into these directions, too.
\\

The author would like to thank T.-S.~H.~Lee, S.~X.~Nakamura, and T.~Sato for their collaborations.
This work was supported by the JSPS KAKENHI Grant Number JP25800149.


\begin{thebibliography}{99}

\bibitem{chicago} 
H.~L.~Anderson, E.~Fermi, E.~A.~Long, and D.~E.~Nagle:
Phys.\ Rev.\  {\bf 85}, 936 (1952).

\bibitem{pdg14}
K.~A.~Olive {\it et al.} (Particle Data Group):
Chin.\ Phys.\ C {\bf 38}, 090001 (2014).

\bibitem{said06}
R.~A.~Arndt, W.~J.~Briscoe, I.~I.~Strakovsky, and R.~L.~Workman:
Phys.\ Rev.\ C {\bf 74}, 045205 (2006).

\bibitem{cqm}
See, e.g., S.~Capstick and W.~Roberts:
Prog.\ Part.\ Nucl.\ Phys.\  {\bf 45}, S241 (2000).

\bibitem{dse}
See, e.g., 
A.~Bashir, L.~Chang, I.~C.~Cloet, B.~El-Bennich, Y.~X.~Liu, C.~D.~Roberts, and P.~C.~Tandy:
Commun.\ Theor.\ Phys.\  {\bf 58}, 79 (2012);
G.~Eichmann, H.~Sanchis-Alepuz, R.~Williams, R.~Alkofer, and C.~S.~Fischer:
Prog.\ Part.\ Nucl.\ Phys.\  {\bf 91}, 1 (2016).

\bibitem{msl07}
A.~Matsuyama, T.~Sato, and T.-S.~H.~Lee:
Phys. Rep. {\bf 439}, 193 (2007).

\bibitem{knls13}
H.~Kamano, S.~X.~Nakamura, T.-S.~H.~Lee, and T.~Sato:
Phys. Rev. C {\bf 88}, 035209 (2013).

\bibitem{knls16}
H.~Kamano, S.~X.~Nakamura, T.-S.~H.~Lee, and T.~Sato:
Phys. Rev. C {\bf 94}, 015201 (2016).

\bibitem{nstar2015}
See, e.g., 
{\it Proceedings of the 10th International Workshop on the Physics of Excited Nucleons (NSTAR2015)},
edited by A. Hosaka {\it et al.}, published as JPS Conf. Proc., Vol 10 (2016).

\bibitem{gw}
M.~L.~Goldberger and K.~M.~Watson, 
{\it Collision Theory}
(John Wiley \& Sons, New York, 1964).

\bibitem{sjklms10}
N.~Suzuki, B.~Juli\'a-D\'iaz, H.~Kamano, T.-S.~H.~Lee, A.~Matsuyama, and T.~Sato:
Phys. Rev. Lett. {\bf 104}, 042302 (2010).

\bibitem{jlms07}
B.~Juli\'a-D\'iaz, T.-S.~H.~Lee, A.~Matsuyama, and T.~Sato:
Phys. Rev. C {\bf 76}, 065201 (2007).

\bibitem{eden}
R.~J.~Eden and J.~R.~Taylor:
Phys. Rev. Lett. {\bf 11}, 516 (1963);
Phys. Rev. {\bf 133}, B1575 (1964).

\bibitem{arndt85}
R.~A.~Arndt, J.~M.~Ford, and L.~D.~Roper:
Phys. Rev. D {\bf 32}, 1085 (1985).

\bibitem{cw90}
R.~E.~Cutkosky and S.~Wang:
Phys. Rev. D {\bf 42}, 235 (1990).

\bibitem{doring09}
M.~D\"oring, C.~Hanhart, F.~Huang, S.~Krewald, and U-G.~Mei{\ss}ner:
Nucl. Phys. {\bf A829}, 170 (2009).

\bibitem{sl2}
B.~Juli\'a-D\'iaz, T.-S. H. Lee, T.~Sato, and L.~C.~Smith:
Phys. Rev. C {\bf 75}, 015205 (2007).

\bibitem{az13}
I.~G.~Aznauryan {\it et al.}: 
Int. J. Mod. Phys. E {\bf 22}, 1330015 (2013).

\bibitem{juelich}
D.~R\"onchen, M.~D\"oring, H.~Haberzettl, J.~Haidenbauer, U.-G.~Mei{\ss}ner, and K.~Nakayama:
Eur. Phys. J. A {\bf 51}, 70 (2015).

\bibitem{bg} 
A.~V.~Anisovich, R.~Beck, E.~Klempt, V.~A.~Nikonov, A.~V.~Sarantsev, and U.~Thoma: 
Eur. Phys. J. A {\bf 48}, 15 (2012);
E.~Gutz {\it et al.} (The CBELSA/TAPS Collaboration):
Eur. Phys. J. A {\bf 50}, 74 (2014).

\bibitem{e45}
H.~Sako and K.~Hicks {\it et al.} (J-PARC E45 Experiment): 
{\it 3-Body Hadronic Reactions for New Aspects of Baryon Spectroscopy},
\url{http://j-parc.jp/researcher/Hadron/en/pac_1207/pdf/P45_2012-3.pdf} .

\bibitem{pi2pi1}
H.~Kamano, B.~Juli\'a-D\'iaz, T.-S.~H.~Lee, A.~Matsuyama, and T.~Sato:
Phys. Rev. C {\bf 79}, 025206 (2009).

\bibitem{pi2pi2}
H.~Kamano: 
Phys. Rev. C {\bf 88}, 045203 (2013).

\bibitem{clas12-proposals}
R.~Gothe {\it et al.}:
Nucleon Resonance Studies with CLAS12 (JLab E12-09-003),
\url{https://www.jlab.org/exp_prog/proposals/09/PR12-09-003.pdf};
D.~S.~Carman {\it el al.}:
Exclusive $N^*\to KY$ Studies with CLAS12 (JLab E12-06-108A),
\url{https://www.jlab.org/exp_prog/proposals/14/E12-06-108A.pdf} .

\bibitem{nstar-kamano}
H.~Kamano,
JPS Conf. Proc. {\bf 10}, 010002 (2016).

\bibitem{joo-smith}
K.~Joo and L.~C.~Smith: private communication.

\bibitem{clas}
K.~Joo {\it et al.} (CLAS Collaboration): 
Phys. Rev. Lett. {\bf 88}, 122001 (2002);
Phys. Rev. C {\bf 68}, 032201 (2003);
H.~Egiyan {\it et al.} (CLAS Collaboration): 
Phys. Rev. C {\bf 73}, 025204 (2006);
K.~Park {\it et al.} (The CLAS Collaboration): 
Phys. Rev. C {\bf 77}, 015208 (2008);
Phys. Rev. C {\bf 91}, 045203 (2015);
M.~Ungaro {\it et al.} (The CLAS Collaboration): 
Phys. Rev. Lett. {\bf 97}, 112003 (2006);
L.~C.~Smith {\it et al.} (CLAS Collaboration): 
AIP Conf. Proc. {\bf 904}, 222 (2007);
I.~G.~Aznauryan {\it et al.} (CLAS Collaboration): 
Phys. Rev. C {\bf 80}, 055203 (2009).

\bibitem{jklmss09}
B.~Juli\'a-D\'iaz, H. Kamano, T.-S.~H.~Lee, A.~Matsuyama, T.~Sato, and N.~Suzuki:
Phys. Rev. C {\bf 80}, 025207 (2009).

\bibitem{ssl2}
N.~Suzuki, T.~Sato, and T.-S.~H.~Lee:
Phys. Rev. C {\bf 82} 045206 (2010).

\bibitem{footnote}
The channel space and number of $P_{33}$ bare $N^*$ states considered in each model
are:
($\pi N$, $\pi \pi N$, $\eta N$, $K\Lambda$, $K\Sigma$)
and two bare $N^*$ states for the current ANL-Osaka analysis;
($\pi N$, $\pi \pi N$, $\eta N$) and two bare $N^*$ states 
for our previous analysis~\cite{jklmss09,ssl2};
and the $\pi N$ channel and one bare $N^*$ state for the Sato-Lee model~\cite{sl2}.

\bibitem{knlskp1}
H.~Kamano, S.~X.~Nakamura, T.-S.~H.~Lee, and T.~Sato:
Phys. Rev. C {\bf 90}, 065204 (2014).

\bibitem{knlskp2}
H.~Kamano, S.~X.~Nakamura, T.-S.~H.~Lee, and T.~Sato:
Phys. Rev. C {\bf 92}, 025205 (2015).

\bibitem{kl16}
H.~Kamano and T.-S.~H.~Lee:
arXiv:1608.03470.

\bibitem{meson1}
H.~Kamano, S.~X.~Nakamura, T.~S.~H.~Lee, and T.~Sato:
Phys.\ Rev.\ D {\bf 84}, 114019 (2011).

\bibitem{meson2}
S.~X.~Nakamura, H.~Kamano, T.~S.~H.~Lee, and T.~Sato:
Phys.\ Rev.\ D {\bf 86}, 114012 (2012).

\bibitem{neutrino1}
H.~Kamano, S.~X.~Nakamura, T.-S.~H.~Lee, and T.~Sato:
Phys.\ Rev.\ D {\bf 86}, 097503 (2012).

\bibitem{neutrino2}
S.~X.~Nakamura, H.~Kamano, and T.~Sato:
Phys.\ Rev.\ D {\bf 92}, 074024 (2015).


\end{thebibliography}
\end{document}